\documentstyle[12pt,epsf]{article}
\begin{document}
\begin{center}
{\Large {\bf LEPTON SCATTERING ON NUCLEI AT {\boldmath $x>1$} AND THE NUCLEON 
SPECTRAL FUNCTION}}
\end{center}

\vspace{0.5cm}

\begin{center}
{\Large {P. Fern\'andez de C\'ordoba, E. Marco, H. M\"uther, E. Oset and 
A. Faessler}}
\end{center}

\vspace{0.3cm}

{\small {\it
Institut f\"ur Theoretische Physic, Universit\"at T\"ubingen, 
72076 T\"ubingen, Germany
}}

\vspace{0.6cm}

\begin{abstract}
{\small{ We analyze the processes of deep inelastic and quasielastic
scattering in the region of $x>1$. These processes are found to be 
very sensitive to the information contained in the nucleon
spectral function in nuclei, particularly to the correlations between
energy and momentum. Calculations are done using two spectral 
functions for nuclear matter and one for finite nuclei, $^{16}$O.
Several approximations are also analyzed and are shown to be 
inaccurate in this region. These results stress the fact that 
the region of $x>1$ contains important information on nuclear 
dynamical correlations.}}
\end{abstract}

\section{Introduction}

Deep inelastic scattering (DIS) and quasielastic scattering (QES)
on nuclei at $x>1$ (being $x$ the Bjorken variable) are a very
interesting tool to explore nuclear dynamics. In DIS on a free
nucleon at rest, the possible values of $x$ are limited,
$0<x<1$, and in QES the value is fixed, $x=1$. DIS and QES on
a nucleus at $x>1$ are possible due to the Fermi motion of the 
nucleons and the interactions between them.

After the discovery of the EMC effect \cite{emc}, the raise
of the ratio of the nuclear structure function to the free one
at $x$ close to 1 was soon identified \cite{bodek} as the effect of the Fermi 
motion of the nucleons. This motivated people to investigate
the region of $x>1$ [2--5], noting that a non vanishing value 
of the structure function for $x>1+ k_{\mbox{\scriptsize F}}/M$
(with $k_{\mbox{\scriptsize F}}$ the Fermi momentum), would require a tail in
the momentum distribution, $n(\vec{p}\,)$, for $k>k_{\mbox{\scriptsize F}}$.
This is indeed the case when the effects of a residual $NN$ interaction
are considered, leading to a correlated many-body system of fermions.

However, it is very dangerous to 
use only the momentum distributions, because in this region it 
is essential to take into account the correlations between energy 
and momentum, which are described by the spectral functions.

In our work \cite{xgt1}, we have used two different spectral
functions for nuclear matter and one for $^{16}$O in order 
to quantify the uncertainties of the many-body approach. We
have used the theoretical framework developed in Ref.~\cite{xlt1},
in which nucleons are treated in a relativistic formalism. We have
also analyzed several approximations widely used in the 
calculations of the EMC effect.

We compare our results with the high $Q^2$ data ($Q^2>60\,$ GeV$^2$)
of BCDMS \cite{12Cdata}, in which DIS dominates and with the low $Q^2$
data ($Q^2\sim 1.3\,$ GeV$^2$) obtained at SLAC \cite{filip},
in which QES is very important.

\section{Formalism for inelastic lepton scattering}
In our work in Ref.~\cite{xlt1} , it is shown that using the relativistic
formalism we can write the hadronic tensor for a nucleus as 

\begin{equation}
W'^{\mu \nu}_A = 4 \int \, d^3 r \, \int \frac{d^3 p}{(2 \pi)^3} \, 
\frac{M}{E (\vec{p})} \, \int^{\mu}_{- \infty} d p^0 S_h (p^0, p) \,
W'^{\mu \nu}_N (p, q)\,,
\end{equation}
where $q$ is the virtual photon momentum, $W'^{\mu \nu}_{N}(p,q)$ is
the average of the hadronic tensor for protons and neutrons
and $S_h (p^0, p)$ is the 
relativistic hole spectral function, normalized as

\begin{equation}
4 \int d^3 r \int \frac{d^3 p}{(2 \pi)^3} \int^{\mu}_{- \infty}
d \omega \; S_h \left(\omega, p;\rho(r)\right) =  A
\end{equation}
in symmetric nuclear matter, where $\mu$ is the chemical potential.
Had we used a nonrelativistic formalism, we would have obtained an 
expression similar to (1) but without the factor $M/E(\vec{p} \,)$
and evaluated using a nonrelativistic spectral function.

Gauge invariance imposes the following structure of the hadronic tensor
in terms of two invariant structure functions $W_{1}$, $W_{2}$,

\begin{equation}
W'^{\mu \nu} = 
\left( \frac{q^{\mu} q^{\nu}}{q^2} - g^{\mu \nu} \right) \;
W_1 + \left( p^{\mu} - \frac{p . q}{q^2} \; q^{\mu} \right)
\left( p^{\nu} - \frac{p . q}{q^2} \; q^{\nu} \right)
\frac{W_2}{M^2}\,.
\label{eq:gauge}
\end{equation}
Since it is customary to show the experimental data for $W_{2}$, we use
Eqs.~(1) and (3) to write $W_{2A}$ in terms of $W_{2N}$
by eliminating $W_{1}$. This is easily accomplished by using 
the expressions for $W'^{xx}$ and $W'^{zz}$ and taking $q$ in the
$z$ direction, as usually done. We get

$$
W_{2A} = - \frac{q^{2}}{|\vec{q}\,|^{2}} \sum_{p,n}
2 \int \, d^3 r \, \int \frac{d^3 p}{(2 \pi)^3} \, 
\frac{M}{E (\vec{p}\,)} \, \int^{\mu}_{- \infty} d p^0 S_h (p^0, p)
$$
\begin{equation}
\times \left[(p^{x})^{2} - \frac{q^{2}}{q}(p^{z} - \frac{p \cdot q}{q^{2}}
|\vec{q}\,|)^{2} \right] \frac{W_{2N}(p,q)}{M^{2}}
\label{eq:w2in}\,,
\end{equation}
where we have substituted a factor 2 of isospin in Eq.~(1) by the explicit
sum over protons and neutrons.

In the Bjorken limit, $q^{0} \rightarrow \infty$, 
$-q^{2}\rightarrow \infty$, we define 

\begin{equation}
x=\frac{-q^{2}}{2M q^{0}} ; \quad x_{N}=\frac{-q^{2}}{2p \cdot q}
\end{equation}
and we find 
\begin{equation}
F_{2A} (x) = \sum_{p,n} 2 \int d^3 r \int \frac{d^3 p}{(2 \pi)^3} 
\frac{M}{E (\vec{p})}
\int_{- \infty}^{\mu} d \omega \, S_h (\omega,p) \frac{x}{x_N}
F_{2N} (x_N)\,,
\label{eq:strucr}
\end{equation}
where $F_{2N}(x)$ is the nucleon structure function (which depends smoothly
on $Q^{2}$). This is the expression found in \cite{xlt1} for the 
$F_{2A}(x)$ structure function in the Bjorken limit.

For the quasielastic contribution to the structure function, $W_{2A}^{Q}$.
Following the steps of Section~3 of Ref.~\cite{xlt1} we can write

$$
W'^{\mu \nu}_A = \sum_{n, p} 2 \int d^3 r \; \int \frac{d^3 p}{(2 \pi)^3} \;
\frac{M}{E (\vec{p})} \; \int ^{\mu}_{- \infty} \; S_h (p^0, p) \, d p^0
$$

\begin{equation}
\frac{M}{E (\vec{p} + \vec{q})} \, \bar{\sum_{s_i}} \, \sum_{s_f} \, 
\langle p+q | J^{\mu} | p \rangle \langle p | J^{\nu} | p+q \rangle^*
\delta (q^{0} + p^{0} - E (\vec{p} + \vec{q}))
\label{eq:wqua}\,,
\end{equation}
where
\begin{equation}
\langle p' |J^{\mu}|p \rangle = \bar u(\vec{p}') 
\left[F_{1}(q) \gamma^{\mu} + i \frac{F_{2}(q)}{2M} 
\sigma^{\mu \nu} q_{\nu}   \right] u(p)\,.
\end{equation}
$W^{Q}_{2A}$ is given by means of Eq.~(\ref{eq:gauge}) for the nucleus
at rest by
\begin{equation}
W^{Q}_{2A} = \left(\frac{q^{2}}{\vec{q}^{2}}\right)^{2} W'^{00}_{A,Q}
- \frac{q^{2}}{\vec{q}^{2}} W'^{xx}_{A,Q}\,.
\label{eq:w2qua}
\end{equation}
By means of Eqs.~(\ref{eq:wqua}) and (\ref{eq:w2qua}) 
we can evaluate the quasielastic structure function $W^{Q}_{2A}$,
and express it using the Sachs form factors $G_E$, $G_M$ \cite{amal},
defined as
$$
G_{E}(q) = F_{1}(q) + \frac{q^{2}}{4M^{2}} F_{2}(q)\,,
$$
\begin{equation}
G_{M}(q) = F_{1}(q) +  F_{2}(q)\,. 
\end{equation}

\section{The nucleon spectral function in nuclear matter and 
         finite nuclei}

We have used three different approaches to evaluate the spectral
function: two evaluated for nuclear matter and used by means of
the local density approximation and one for the nucleus of $^{16}$O.

{\it 3.1. Semiphenomenological approach in nuclear matter}

This model is described in detail in Ref.~\cite{semi}. It 
evaluates the nucleon self-energy diagrams in a nuclear medium
but uses the input from the $NN$ experimental cross section and
the polarization of the $NN$ interaction to circumvent the use of the 
$NN$ potential and the ladder sums. In Ref.~\cite{xlt1} it is explained
how to include relativistic effects in this spectral function.

{\it 3.2. Microscopic approach in nuclear matter}

This spectral function has been evaluated using the techniques described 
in Ref.~\cite{micmat}. The starting point of this many-body
calculation is a Brueckner-Hartree-Fock calculation of nuclear matter
considering the realistic OBE potential 
for the $NN$ interaction.

{\it 3.3. Microscopic approach for finite nuclei}

The spectral function can be calculated directly for finite nuclei
using the procedure described and applied in Ref.~\cite{micfin}. The
nucleon spectral function is separated in two parts: the quasiparticle 
contribution to the spectral function around
the quasiparticle pole (the occupied states of the Shell model)
and the background contribution which contains the information about 
the spectral function at energies away from the respective 
quasiparticle pole. This background contribution,
$S_{h}^B (\omega,p;\rho(r))$, is evaluated for
nuclear matter,

\begin{equation}
S_{h,A}(\omega, p) = S^{QP}(\omega,p) + 4 \int d^3r S_{h}^B
(\omega,p;\rho(r))\,.
\label{eq:sfin}
\end{equation}

\newpage
{\it 3.4. Approximations to be avoided}

We have also used several approximations in order to see their accuracy 
in this region.

i) Non interacting Fermi sea.

  The expression used for the spectral function in this approximation is

\begin{equation}
S^{UFS}_{h} (\omega, p;\rho )  = n_{FS} (\vec{p}) \delta (\omega
 - E (\vec{p}) - \Sigma)\,,
\label{eq:nufs}
\end{equation}
where $n_{FS} (\vec{p})$ is the occupation number, 0 for momenta above the
local Fermi momentum and 1 for momenta below the local Fermi momentum.
$\Sigma$ is the self-energy which accounts for the binding of the nucleons.

ii) Momentum distribution

Since large momentum components are needed to generate $F_{2A} (x)$
at $x>1$, one is tempted to use Eq.~(\ref{eq:nufs}) but using the realistic
momentum distribution as an improvement over the previous approximation.
The momentum distribution is given by

\begin{equation}
n_{I} (\vec{p}) = \int_{- \infty}^\mu S_h (\omega, p) d \omega\,.
\label{eq:ndei}
\end{equation}

iii) Momentum distribution and the corresponding mean value of
the energy

Finally we consider an approximation in which we take for
the spectral function the momentum 
distribution of Eq.~(\ref{eq:ndei}) but for the value of 
the energy in the $\delta$
function we take the mean value of the energy,

\begin{equation}
S_{h}^{MED} (\omega,p;\rho ) = n_{I} (\vec{p}\,)  \delta (\omega - 
\langle \omega (\vec{p})\rangle)\,.
\label{eq:swmed}
\end{equation}
In Fig.~1 we can see 
the mean value of the energy as a function of $|\vec{p} \,|$.
This figure shows that there is an important correlation
between the momenta and the mean value of the energy for the bound
nucleons. All approximations of Sect.~{\it 3.4} lack this correlation
and it will be shown that they give wrong results because of that.

\medskip
\begin{figure}[h]
\hspace{1.0in}
\epsfxsize=3.5in
\centerline{\epsffile{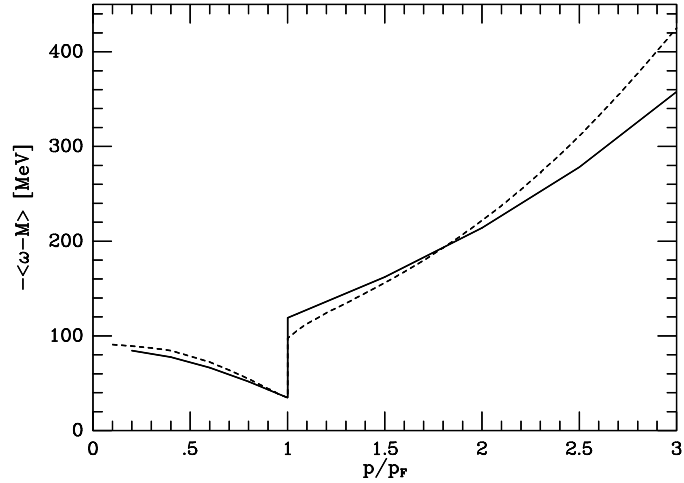}}
\caption{Mean value of the energy of the nucleons as a function of 
$|\vec{p}\,|$, at $\rho = \rho_{0}$. Solid line: microscopic model 
\protect\cite{micmat}; dashed line: semiphenomenological model 
\protect\cite{semi}.}
\label{fig1}
\end{figure}
\medskip

\section{Results}
\medskip
\begin{figure}[t]
\hspace{1.0in}
\epsfxsize=3.5in
\centerline{\epsffile{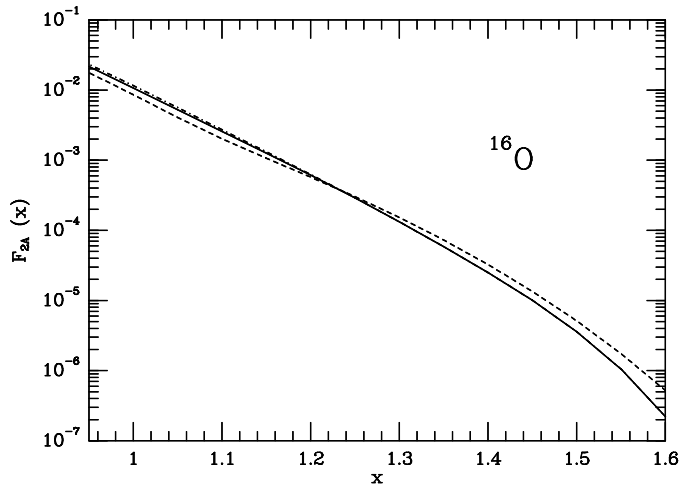}}
\caption{Results obtained for the structure function of $^{16}$O. 
Solid line: microscopic nuclear matter model \protect\cite{micmat}; 
dot-dashed line: microscopic finite nuclei model \protect\cite{micfin}; 
dashed line: nonrelativistic semiphenomenological model \protect\cite{semi}.
}
\label{fig2}
\end{figure}
\medskip
In Fig.~2 we show the results for $F_{2A} (x)$ calculated with the three
different spectral functions introduced in Sects.~{\it 3.1--3.3},
for the case of $^{16}$O. Since the microscopic nuclear matter 
and finite nuclei approaches are nonrelativistic, we have also taken
the nonrelativistic version of the semiphenomenological approach.
The experimental values for $F_{2N} (x)$ are taken from 
Ref.~\cite{f2}. The results obtained with the two spectral functions
for nuclear matter (solid line and dashed line) are rather similar.
At $x\simeq 1$ the microscopic spectral function provides results
about 20~\% higher than the semiphenomenological one. At values
of $x\simeq 1.22$ the two approaches coincide and for $x\simeq1.5$,
the semiphenomenological approach provides values of $F_{2A(x)}$
about 40~\% larger than the microscopic one.

The results obtained with the spectral function for finite nucleus,
Sect.~{\it 3.3}, are represented by the dot-dashed line in Fig.~2.
They should be compared with those displayed by the solid line
since the background contribution to Eq.~(\ref{eq:sfin})
is obtained from the same nuclear matter result. At $x\simeq 1$
the results with the spectral function of the finite nucleus
are about 8~\% higher than with the nuclear matter approach.
The differences become smaller as $x$ increases and for values of
$x\simeq 1.5$ the two approaches give the same results. This latter
fact is telling us that at large values of $x$ one is getting 
practically all contributions from the background part of the spectral
function and none from the quasiparticle part.

\medskip
\begin{figure}
\hspace{1.0in}
\epsfxsize=3.5in
\centerline{\epsffile{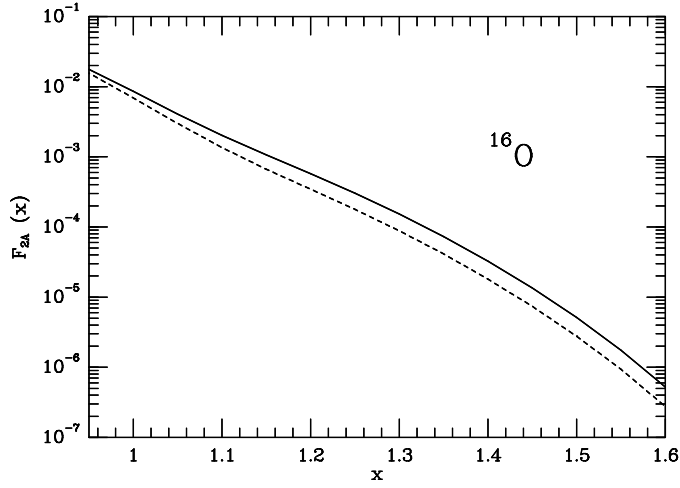}}
\caption{Results obtained for the structure function of $^{16}$O using
the semiphenomenological model \protect\cite{semi}. Solid line: 
nonrelativistic formalism; dashed line: relativistic formalism.
}
\label{fig3}
\end{figure}
\medskip

In Fig.~3 we show the results obtained with the semiphenomenological
approach using the relativistic and nonrelativistic formalisms. The
relativistic corrections induce a reduction of 25~\% around $x=1$
and roughly reduce the structure function $F_{2A}(x)$
to one half of the nonrelativistic results at $x\simeq 1.5$. The effects
are so important because we get the contribution from the components
of high momentum. 

\medskip
\begin{figure}
\hspace{1.0in}
\epsfxsize=3.5in
\centerline{\epsffile{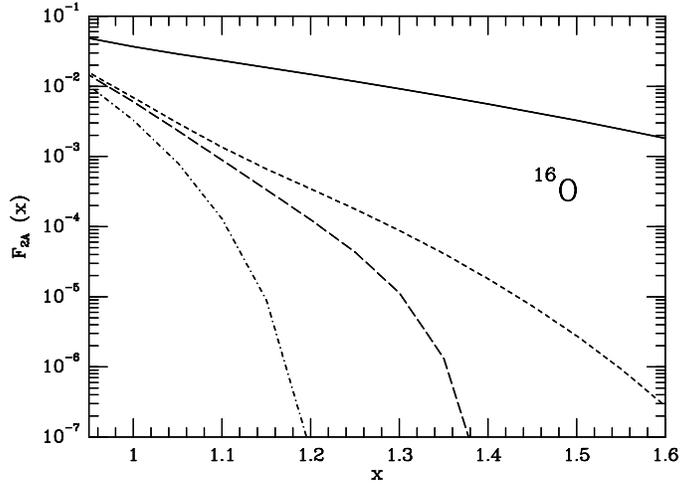}}
\caption{Results obtained for the structure function of $^{16}$O using 
different approximations. Dot-dashed line: non interacting Fermi sea, 
Eq.~(\protect\ref{eq:nufs});  solid line: momentum distribution, 
Eq.~(\protect\ref{eq:ndei}); long-dashed line: momentum distribution 
of the correlated Fermi sea and average energy 
$\langle \omega (\vec{p}\,)\rangle$, Eq.~(\protect\ref{eq:swmed}); 
short-dashed line:
spectral function, \protect\cite{semi}. $Q^2 =$ 5  GeV$^2$.
}
\label{fig4}
\end{figure}
\medskip

Results obtained for $F_{2A} (x)$ using the different approximations
discussed in Sect.~{\it 3.4} are displayed in Fig.~4. The dot-dashed line
represents the non interacting Fermi sea, Eq.~(\ref{eq:nufs}). We
can see that at $x\simeq 1$ it already provides a structure function 
of around a factor two smaller than the one obtained with the proper
spectral function (short-dashed line). At higher values of $x$,
the differences are bigger, showing clearly that one is exploring the
region of large momenta, above the Fermi momentum, which are 
not accounted for by the non interacting Fermi sea.
Another approximation corresponds to using the realistic momentum 
distribution $n_{I} (\vec{p})$ of Eq.~(\ref{eq:ndei}) and associating an
energy to each $\vec{p}$ given by its kinetic energy plus a potential,
Sect.~{\it 3.4} ii). 
The results (solid line) are outrageously wrong. This
demonstrates that the naive use of a momentum distribution, although
calculated in a realistic way, may lead to results which are worse than
those obtained for an uncorrelated system, if one does not treat the
energy-momentum correlation properly. The results obtained using
the momentum distribution and the mean value of the energy
are shown in the curve with long dashes. This is the better 
approximation, although the discrepancies with the exact results 
are still large enough to discourage it too.

\medskip
\begin{figure}
\hspace{1.0in}
\epsfxsize=3.5in
\centerline{\epsffile{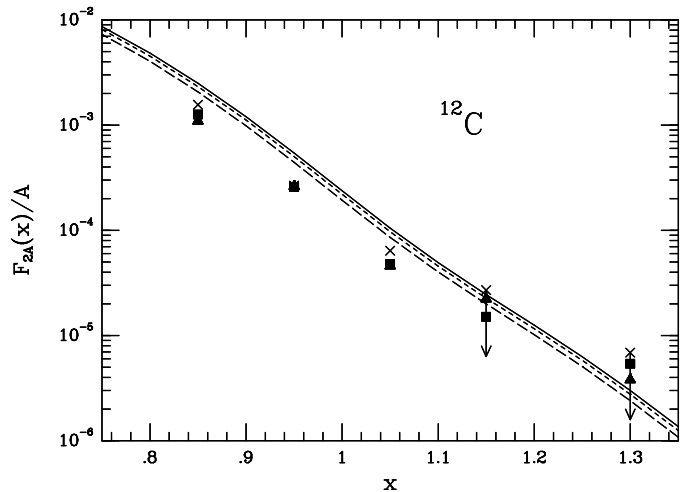}}
\caption{Results for the structure function of $^{12}$C at $Q^2 = 61, 85 $
and $150$  GeV$^2$ (solid, short-dashed and long dashed lines
respectively). The data are from Ref.~\protect\cite{12Cdata} 
crosses for 61 GeV$^2$, squares for 85 GeV$^2$ and triangles 
for 150 GeV$^2$. The data for the two largest
values of $x$ are upper bounds.
}
\label{fig5}
\end{figure}
\medskip

In Fig.~5 we see the comparison of our predictions (using the
$Q^{2}$ dependent structure function of \cite{duke}) to the
measurements done at $Q^{2}=61, 85$ and 150 GeV$^{2}$ in 
Ref.~\cite{12Cdata}. The agreement with the data is qualitative, 
the slope is well reproduced but the theoretical results are in
average 40~\% higher than experiment up to $x \simeq 1.05$. At
$x\simeq 1.15$ and $1.3$ there are only upper bounds which
are compatible with our predictions.

\medskip
\begin{figure}
\hspace{1.0in}
\epsfxsize=3.5in
\centerline{\epsfbox{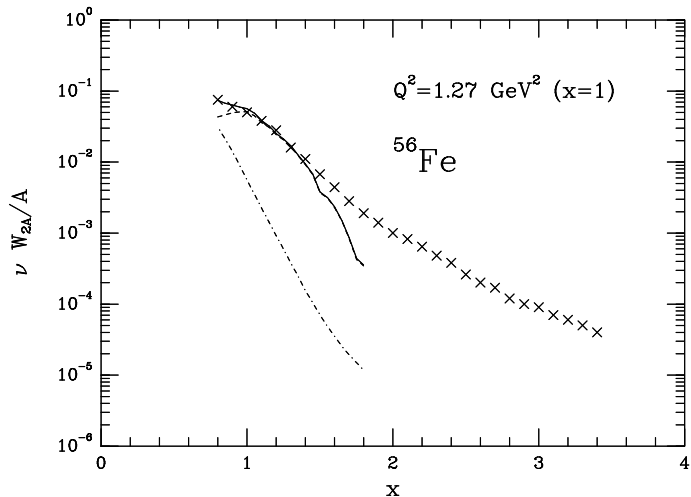}}
\caption{Results for the structure function for $^{56}$Fe at $Q^2 = 1.27$
GeV$^2$ when $x=1$. Dot-dashed line: inelastic contribution; 
dashed line: quasielastic contribution; solid line: 
total contribution.
}
\label{fig6}
\end{figure}
\medskip

In Fig.~6 we can see the results obtained at low values of $Q^{2}$
($Q^{2} = 1.27$ GeV$^{2}$ at $x=1$). In this region we include
the quasielastic contribution of Eq.~(\ref{eq:w2qua}) and the inelastic
contribution, Eq.~(\ref{eq:w2in}). For low values of $Q^2$ we use
a parametrization of the structure functions \cite{bodek, stein}
where there is a part corresponding to the excitation of the low lying
resonances (usually called the inelastic part) and a smooth part 
for the excitation in the continuum which would stand for the deep
inelastic part. We can see that the quasielastic contribution is 
dominant in all the range of the figure and peaks around
$x=1$. The spread of the quasielastic contribution is due to Fermi 
motion and binding. The inelastic contribution is small in all the range
of the figure compared to the quasielastic contribution. However,
at values of $x<1$ the inelastic contribution dominates. For values
of $x>1$ the strength of the structure function is completely
dominated by the quasielastic contribution. The agreement with 
the experiment is rather good up to values of $x$ around $1.4$,
from there on our results start diverging from the data.

\section{Conclusions}
We have investigated the region of $x>1$ and shown that it provides
information on the dynamical properties of nuclei beyond the
approximate shell model structure. In order to study these
correlations it is important to go to high values of $Q^{2}$,
since at low values the quasielastic contribution dominates
and it is not so sensitive to the nuclear correlations.

We have seen that in order to treat correctly these correlations,
it is important to use spectral functions, without indulging
in any approximation. Moreover, since the contribution in this
region comes from high momentum components, relativistic
effects should be taken into account in the spectral function.

\end{document}